\documentclass[fleqn,usenatbib]{mnras}

\usepackage[T1]{fontenc}

\DeclareRobustCommand{\VAN}[3]{#2}
\let\VANthebibliography\thebibliography
\def\thebibliography{\DeclareRobustCommand{\VAN}[3]{##3}\VANthebibliography}

\usepackage{graphicx}	
\usepackage{amsmath}	
\usepackage{amssymb}	

\usepackage{newtxtext,newtxmath}

\newcommand       \RV           {{R_V}}
\newcommand       \AV           {{A_V}}

\newcommand       \mum         {\rm \mu m}
\defcitealias{car89}{CCM89}
\defcitealias{fit99}{F99}

\usepackage{multirow}

\title[Dust models of SN 2010\MakeLowercase{jl}]{Dust Models for the Extinction of Type II\MakeLowercase{n} Supernova SN 2010\MakeLowercase{jl}}

\author[J. Li et al.]{
Jun Li,$^{1,2}$\thanks{E-mail: lijun@mail.bnu.edu.cn (JL)}
Jian Gao,$^{1}$\thanks{E-mail: jiangao@bnu.edu.cn (JG)}
Biwei Jiang$^{1}$\thanks{E-mail: bjiang@bnu.edu.cn (BWJ)}
and Zesen Lin$^{3,4}$
\\
$^{1}$Department of Astronomy, Beijing Normal University, Beijing 100875, P. R. China\\
$^{2}$Department of Astronomy, University of Massachusetts, Amherst, MA 01003, USA\\
$^{3}$Key Laboratory for Research in Galaxies and Cosmology, Department of Astronomy, University of Science and Technology of China, Hefei 230026, China\\
$^{4}$School of Astronomy and Space Sciences, University of Science and Technology of China, Hefei 230026, China
}

\date{Accepted XXX. Received YYY; in original form ZZZ}

\pubyear{2021}

\begin{document}
\label{firstpage}
\pagerange{\pageref{firstpage}--\pageref{lastpage}}
\maketitle

\begin{abstract}
The unusual extinction curves of SN 2010jl provide an excellent opportunity to investigate the properties of dust formed by core-collapse supernovae. By using a series of dust models with different compositions and grain size distributions, we fit the extinction curves of SN 2010jl and find that a silicate-graphite mixture dust model characterized by exponentially cutoff power-law size distributions can well reproduce its unusual extinction curves. The best-fit results show that the extinctions derived from the dust models are consistent with the observed values at all epochs. However, the total-to-selective extinction ratio $\RV$ is about 2.8 - 3.1, which is significantly smaller than the value of $\RV\approx6.4$ derived by Gall et al. The best-fit models indicate that the dust grains around SN 2010jl are possibly composed of small-size astronomical silicate grains and micron-size graphite grains. In addition, by fitting the optical to mid-infrared spectral energy distribution, we find that the dust mass around SN 2010jl increases with time, up to $0.005\,M_{\odot}$ around 1300 days after peak brightness, which is consistent with previous estimates.
\end{abstract}

\begin{keywords}
circumstellar matter - dust, extinction - supernovae: individual (SN 2010jl)
\end{keywords}



\section{Introduction} \label{sec:intro}

 Core-collapse supernovae (CCSNe) originated from massive stars are suggested to be one of the major production sites of cosmic dust, which is supported by the existence of large amount of dust in very young universe \citep{ber03,pri03,rob04,cul17}. In the meantime, supernovae also destroy and alter the properties of the interstellar dust due to strong shock waves. However, it is unclear how much supernovae may contribute to and influence the interstellar dust \citep{dwe06}. 

Type IIn supernovae (SNe IIn) are a particular class of Type II CCSNe, characterized by their narrow ($\sim 100\,{\rm km\,s^{-1}}$) and often intermediate ($\sim 1000\,\rm{km\,s^{-1}}$) spectral lines of hydrogen and helium, which appear along with normal broad ($\sim 10,000\,\rm{km\,s^{-1}}$) lines due to the expansion of the ejecta \citep{sch90}. Some researches found that dust mass in SNe IIn increases continuously after explosion \citep{gal14,bev19}.  \cite{str12} reported that dust mass in type IIn SN 2005ip and SN 2006jd ranges from $10^{-4}$ to $10^{-3}\,M_\odot$ at early times (<1600 days). \cite{bev19} proposed rapid dust formation in SN 2005ip and found that dust mass increases significantly to 0.1$\,M_\odot$ at $\sim$4000 days.
For general type of CCSNe, there have been some controversy about continuous dust formation. Very small dust mass (i.e. $10^{-4}-10^{-2}\,M_\odot$) is observed a few hundred days after explosion \citep[e.g.,][]{woo93,kot09,gal11}, while a large dust mass of $0.1-1\,M_\odot$ can be found in some old supernova remnants \citep[e.g.,][]{bar10,gom12,tem17}. Moreover, it remains unknown how much dust will eventually survive the SN shocks.
A better understanding about the dust physical properties in CCSNe would be helpful to find answers to these questions.

Although there is not yet a general consent about their progenitors, SNe IIn are originated from massive stars, either luminous blue variable, red supergiant or yellow supergiant \citep{dwe17}. The circumstellar envelope of massive stars is also the dust factory, because these massive stars experience very strong stellar wind with extremely high mass-loss rate, i.e. $10^{-4}$ to $ 10^{-2}\,M_{\odot}\,{\rm yr^{-1}}$, in some cases up to $\sim 0.1 \,M_{\odot} \,{\rm yr^{-1}}$, which are orders of magnitude higher than normal Type II progenitors \citep{fox11}. The strong interaction between the SN ejecta and circumstellar envelope materials (CSM) would lead the dense shell to undergoing strong radiative cooling, thus the temperature could decrease to the dust condensation temperature while the density is still high \citep{poz04}. It is estimated that SNe IIn account for more than half of all known SNe with infrared (IR) emission at late phase, which implies the presence of warm dust, i.e. either the pre-existing CSM dust or newly-formed ejecta dust grains \citep{kot05,wil08}.

SN 2010jl was discovered in the interacting galaxy UGC 5189A at a distance of 48.9\,Mpc on 2010 Nov 3.5 by \cite{new10},
and identified as a Type IIn with a peak absolute magnitude of roughly $-\,20\,\rm mag$ by \cite{ben10} and \cite{yam10}. A significant IR excess that indicated the presence of dust around SN 2010jl was discovered no later than 90 days after the explosion \citep{and11}. However, there are debates regarding the origin of this dust, as well as its mass, composition and grain size distribution. Some researches interpret the IR excess as the evidence of newly-formed dust \citep{smi12,mae13,gal14}, while others suggest a pre-existing and unshocked CSM dust grains \citep{and11,fox13,fra14}. Moreover, \cite{bev20} and \cite{chu18} propose that SN 2010jl presents both evidence of pre-existing dust in CSM and newly-formed dust in cold dense shell and/or ejecta.

\cite{gal14} derived the extinction curves of SN 2010jl from the attenuation of emission lines, and found rapid (in 40 - 240 days) dust formation and inferred the presence of very large ($>1$ micron) grains. In order to model their derived extinction curves of SN 2010jl (see Figure 2b in \cite{gal14}), they combined the extinction curves of the Milky Way (MW) or the Small Magellanic Cloud (SMC) with a grey extinction curve which is supposed to contribute about 40\% of the extinction in the $V$ band. With this extinction model, they estimated the total-to-selective extinction ratio $\RV \approx 6.4$ for SN 2010jl. By assuming the grains to be amorphous carbon, they used a Mathis-Rumpl-Nordsieck power-law size distribution \cite[hereafter MRN]{mat77},  to fit their extinction with model parameters of $\sim a^{-3.5}$ [$a_{\rm min}=0.001\,\mum$, $a_{\rm max}=4.2\,\mum$].
By considering the near-infrared (NIR) emission of dust but without mid-infrared (MIR) emission, the dust mass around SN 2010jl derived by \cite{gal14} was about $10^{-4}$ to $10^{-3} \, M_\odot$. However, their model seriously underestimated the extinction in the short wavelength range (i.e. H$\delta$), thus it correspondingly lead to the underestimation of the contribution of small-size dust grains. 

In this work, we aim to understand the composition and size distribution of the dust grains to reproduce the extinction curves of SN 2010jl obtained by \cite{gal14}. We use the dust model composed of mixed silicate and carbonaceous dust grains. Besides MRN size distribution, we also adopt an exponentially cutoff power-law size distribution \cite[hereafter KMH]{kim94} to probe the dust properties around SN 2010jl. In the meantime, with the derived dust properties and the observed optical to MIR emission of SN 2010jl, the evolution of dust mass around SN 2010jl is investigated.

This paper is organized as follows. Section \ref{sec:data} describes the extinction curves and MIR photometry data of SN 2010jl. Then we report our dust models in Section \ref{sec:model}. The best-fit results are presented in Section \ref{sec:result} and discussed in Section \ref{sec:dis}. Finally, we summarize the results in Section \ref{sec:con}.

\section{Data of SN 2010jl} \label{sec:data}

\subsection{Extinction Data}

Except for Type Ia supernovae, it is difficult to determine the extinction curves of other types of supernovae, because they have no constant intrinsic spectral energy distribution (SED) or absolute magnitude at the maximum brightness. By analysing the line profiles of the most prominent hydrogen, helium and oxygen emission lines of SN 2010jl, \cite{gal14} derived the wavelength-dependent attenuation properties of dust at each epoch (from 44 to 239 days past peak brightness) from the ratios of integrated line profiles. They found that the extinctions of SN 2010jl in some specified bands increase with time (see their Figure 2a), which suggests that the amount of dust increases with time. The wavelength range covers $0.4\,\mum-2.2\,\mum$ and the extinction in the $V$ band ($\AV$) increases continuously from 0.1\,mag at 44 days to 0.6\,mag at 239 days past peak brightness.

\begin{figure}
	\includegraphics[width=\columnwidth]{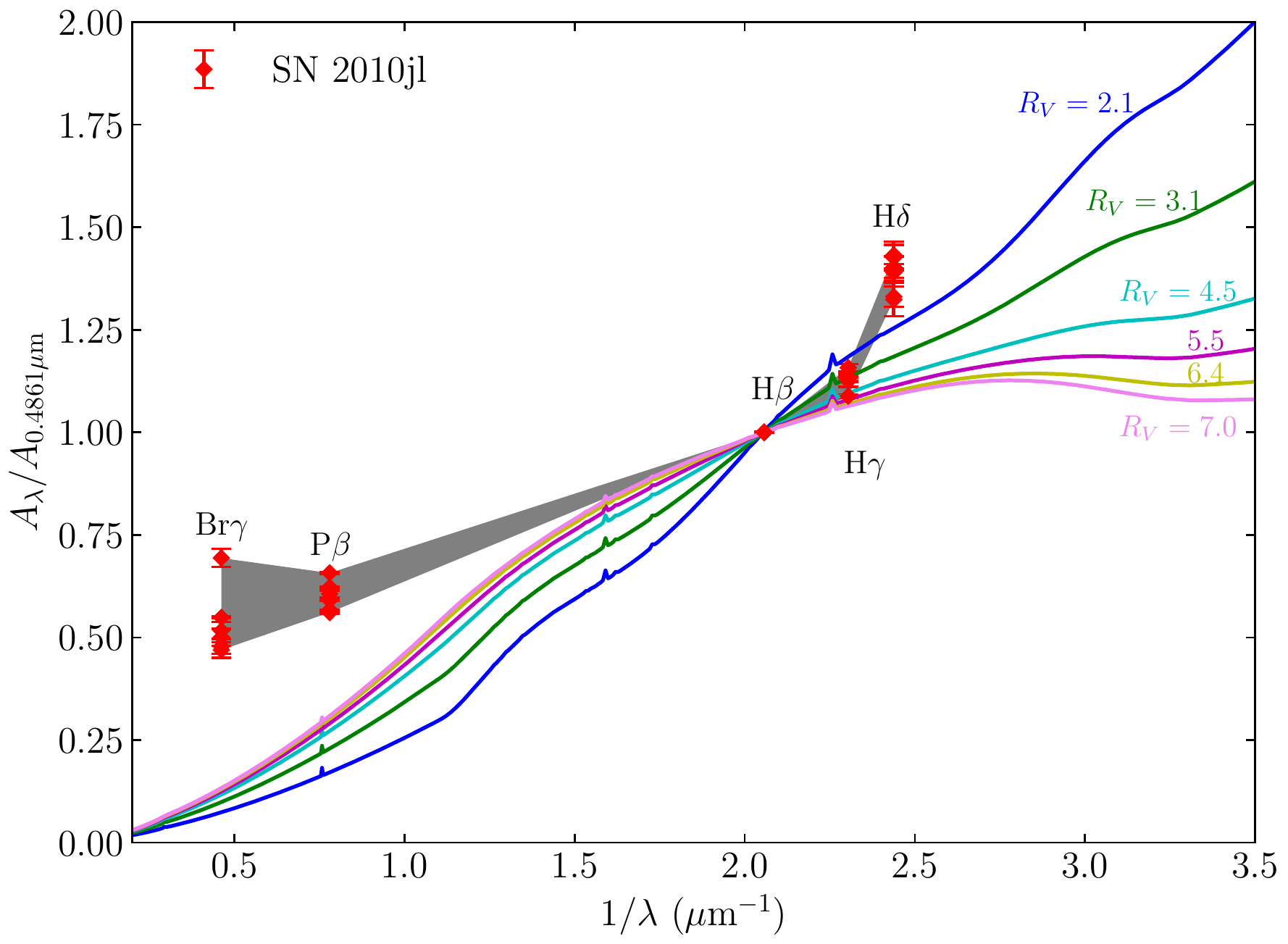}
    \caption{Comparison between the observed extinction curves of SN 2010jl taken from \protect\cite{gal14} and the \citetalias{car89} extinction curves with $\RV$ ranging from $2.1$ to $7.0$. The observed extinction values $A_\lambda/A_{0.4861\mum}$ (normalized to the extinction at $\rm H\beta$) are showed with filled diamonds, while the amplitude of variation of $A_\lambda/A_{0.4861\rm \mu m}$ is represented with grey-shaded area. The \citetalias{car89} extinction curves with different $\RV$ values are showed with different colored solid lines.}
    \label{fig:ccm}
\end{figure}

\cite{gal14} derived the total-to-selective extinction ratio $\RV=\,\AV / E(B-V)\approx6.4$ of SN 2010jl indirectly through their mixed extinction model (grey+MW/SMC extinction model). This $\RV$ value of SN 2010jl is much larger than the average value of the MW (e.g. $\RV \approx 3.1$), even larger than that in the dense region towards HD 36982 (e.g. $\RV \approx 5.8$) (\citealt{car89}, hereafter \citetalias{car89}; \citealt{fit99}, hereafter \citetalias{fit99}). Their extinction laws of SN 2010jl are normalized by the extinction of $\rm H\beta$ at 4861\,\AA\,for all epochs, as shown with diamonds in Figure \ref{fig:ccm}. These extinction laws at different epochs indicate that the properties of dust around SN 2010jl may have small change with time.

Figure \ref{fig:ccm} also illustrates the comparison between the observed extinctions of SN 2010jl and the MW extinction curves of $\RV\,=\,2.1\,-\,7.0$ calculated with the mathematical model provided by \citetalias{car89}. Obviously the extinction curves of SN 2010jl can not be well explained with any \citetalias{car89} models. The long-wavelength side of extinction curves of SN 2010jl are clearly much flatter than \citetalias{car89}'s, while the short-wavelength side looks much steeper at the $\rm H\delta$ wavelength. It can be noted that there are upturns or observed extinction excesses at $\rm H\delta$ presented at all epochs except for at earliest 44 days. Although \cite{gal14} attribute the upturn of observed extinction at $\rm H\delta$ to systematic effects caused by intrinsic line changes, there is no independent evidence to support such a cause. It is intriguing that the intrinsic changes only appear on $\rm H\delta$ lines. By assuming that these upturns are caused by the dust extinction, the physical properties of dust around SN 2010jl are investigated with detailed dust models.

The significant difference between the extinction laws of SN 2010jl with that of the MW suggests that the properties of dust grains around SN 2010jl are quite different with those in the MW. In this work, instead of using the mathematical extinction models, i.e. the \citetalias{car89} model and \citetalias{fit99} model, the detailed physical dust models are used to analyze the extinction curves of SN 2010jl and probe the properties of dust grains around SN 2010jl.

\subsection{MIR Photometry}\label{sec:data_ir}

\begin{table*}
	\centering
	\caption{\emph{Spitzer}/IRAC mid-infrared photometry of SN 2010jl. Figures in parentheses give the uncertainties.}
	\label{tbl:mid-infrared}
	\begin{tabular}{ccccccc}
    \hline
    \hline
    \multirow{2}{*}{MJD} & Epoch$^a$ & \multicolumn{2}{c}{This work} &\multicolumn{2}{c}{Literature} &  \multirow{2}{*}{References} \\
& (days)& $3.6\,\mum$\,($\rm mJy$) & $4.5\,\mum$\,($\rm mJy$) & $3.6\,\mum$\,($\rm mJy$) & $4.5\,\mum$\,($\rm mJy$) & \\
    \hline
    2,455,570 &   82 & 4.65(0.011) & 4.62(0.010)  & 4.04(0.14), 4.14(0.11) & 4.52(0.18), 4.32(0.12) & 1, 2     \\
    2,455,732 &  244 & 4.17(0.013) & 4.49(0.012)  & 5.65(0.17), 3.68(0.10) & 5.80(0.27), 4.18(0.12) & 2, 3     \\
    2,455,740 &  252 & 4.19(0.011) & 4.61(0.010)  & \dots      & \dots      & \dots   \\
    2,455,943 &  455 & 8.91(0.017) & 8.66(0.018)  & 8.55, 11.75(0.31)       & 8.3, 10.84(0.27)        & 2, 4     \\
    2,456,099 &  611 & 9.09(0.018) & 8.99(0.019)  & 8.63, 12.04(0.32)       & 8.66, 11.35(0.27)      & 2, 4     \\
    2,456,322 &  834 & 8.13(0.016) & 8.51(0.018)  & 7.73, 10.62(0.25)  & 8.24, 10.69(0.25)      & 2, 5     \\
    2,456,472 &  984 & 9.56(0.019) & 10.46(0.020) & 9.21, 8.81(0.09)       & 10.29, 9.81(0.10)      & 2, 5     \\
    2,456,476 &  988 & 6.83(0.015) & 7.87(0.016)  & 6.77       & 7.94       & 5     \\
    2,456,846 & 1358 & 5.23(0.014) & 6.64(0.015)  & \dots      & \dots      & \dots \\
    2,456,847 & 1359 & 4.60(0.012) & 5.95(0.013)  & 4.31, 4.67(0.12)       & 5.70, 6.09(0.16)      & 2, 5     \\
    2,457,228 & 1740 & 2.75(0.010) & 4.03(0.010)  & 2.38       & 3.81       & 5     \\
    2,457,811 & 2323 & 1.23(0.006) & 1.83(0.007)  & \dots      & \dots      & \dots \\
    2,458,372 & 2884 & 0.67(0.009) & 0.88(0.009)  & \dots      & \dots      & \dots \\
    \hline
    \multicolumn{7}{@{}l@{}}{$^a$ Relative to peak brightness date, 2010 October 18.6 \protect\citep[MJD = 2,455,488;][]{gal14}.}\\
    \multicolumn{7}{@{}l@{}}{References.~(1) \protect\citet{and11}; (2) \protect\citet{bev20}; (3) \protect\citet{fox13}; (4) \protect\citet{fra14}; (5) \protect\citet{sar18}.}
    \end{tabular}
\end{table*}

\begin{figure}
	\includegraphics[width=\columnwidth]{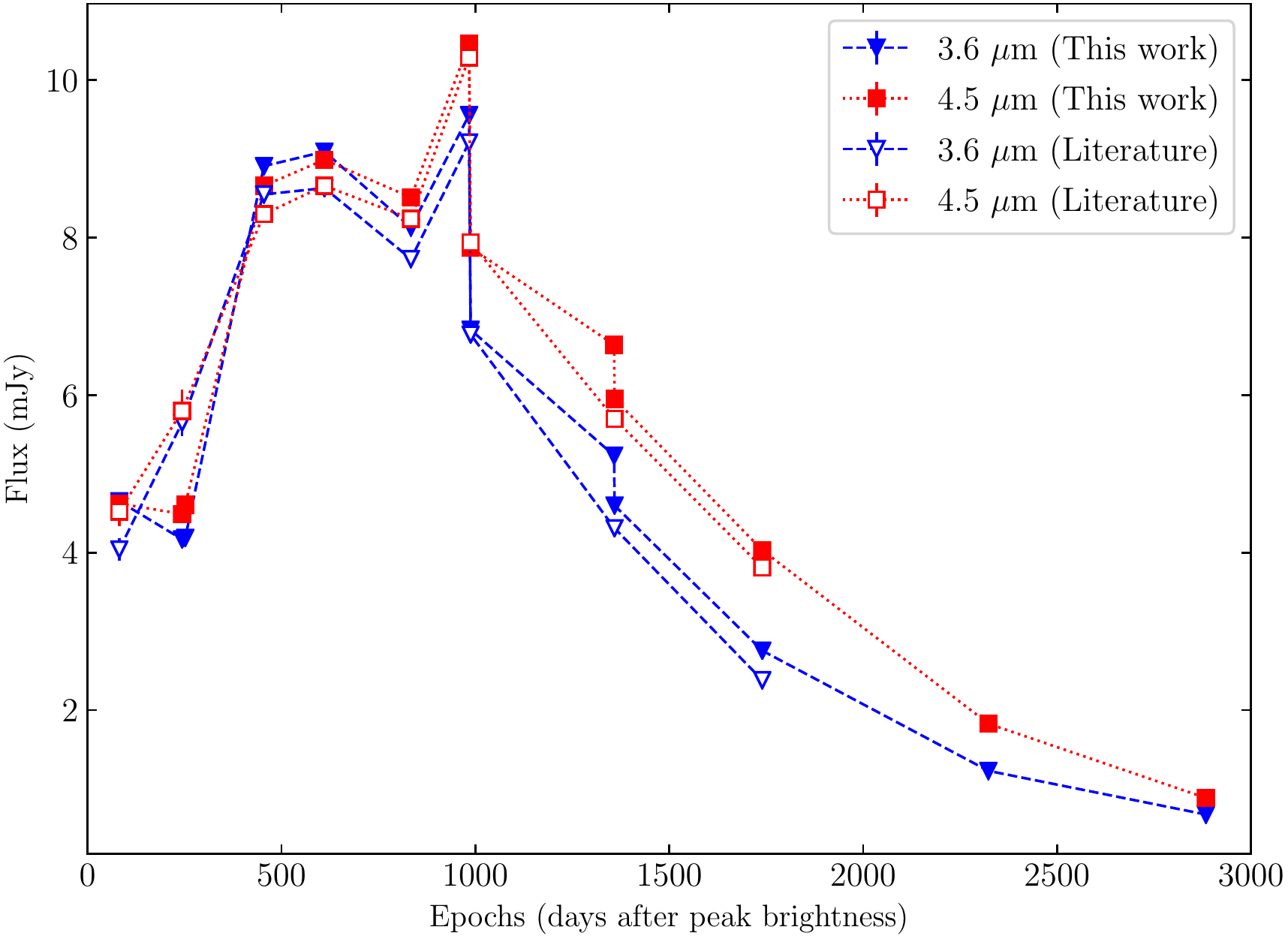}
    \caption{MIR evolution at 3.6\,$\mum$ and 4.5\,$\mum$ of SN 2010jl during the $\simeq$ 4 years after peak brightness. The photometric data extracted in this work are denoted with filled symbols, i.e. filled triangles for 3.6\,$\mum$ and filled squares for 4.5\,$\mum$, respectively. Correspondingly, the data taken from literature are denoted with hollow triangles/squares for comparison. Note that the error bars on the data points are too small to be visible on the figure. All data in this plot are listed in Table \ref{tbl:mid-infrared}.}
    \label{fig:phot}
\end{figure}

In order to estimate the dust mass around SN 2010jl, we carry out a photometric analysis of the entire available \emph{Spitzer}/Infrared Array Camera (IRAC) data of SN 2010jl. All the photometric data of SN 2010jl are only detected in two warm bands, i.e. 3.6\,$\mum$ and 4.5\,$\mum$. The post-basic calibrated data (PBCD) can be accessed from The \emph{Spitzer} Heritage Archive\footnote{https://sha.ipac.caltech.edu/applications/Spitzer/SHA/}. We perform aperture photometry on the PBCD using the \texttt{phot} task in IRAF\footnote{IRAF is distributed by the National Optical Astronomy Observatories, which are operated by the Association of Universities for Research in Astronomy, Inc., under cooperative agreement with the National Science Foundation.}. We also collect the previously published \emph{Spitzer} photometric values before 1740 days, which are generally consistent with the photometric results in this work (see Table \ref{tbl:mid-infrared}). We adopt the fluxes extracted in this work for later analysis. All the data are displayed in Figure \ref{fig:phot}.

These two warm bands of \emph{Spitzer}/IRAC span the peak of blackbody emission from dust with temperature ranging from $500\,\rm K$ to $1000\,\rm K$ \citep{fox11}. By combining the emission of SN 2010jl at $3.6\,\mum$ and $4.5\,\mum$ bands with optical to NIR photometry data taken from \cite{fra14}, the evolution of SEDs offers an opportunity to constrain the masses and temperatures of the dust grains around SN 2010jl. 

Additionally, the Stratospheric Observatory For Infrared Astronomy(\emph{SOFIA}) observation of SN 2010jl was made on 2014 May 5-6 (around 1294 days past peak brightness) by \cite{wil15}. It gives the $1\sigma$ flux upper limit of $4.2$\,mJy at $11.1\,\mum$. Besides, \cite{bev20} reported another observation with VLT/VISIR in which the $3\sigma$ flux upper limit at 10.7$\,\mum$ at 510\,days is 2\,mJy, a non-zero fraction of silicate was speculated in their SEDs prediction. However, both the two flux upper limits can not completely eliminate the presence of silicate grains around SN 2010jl, they provide us an additional constraint on the silicate component in our dust models, which will be discussed in Section \ref{sec:mass_ratio}.

\section{Dust Models} \label{sec:model}
\subsection{Modelling the Extinction Curves}\label{subsec:ext-model}

The dust model used in this work is a mixed dust model consisting of astronomical silicate and graphite or amorphous carbon. The dielectric constants are taken from \cite{dra84} for silicate and graphite, and from \cite{rou91} for amorphous carbon. Correspondingly, the mass densities are $\rho_{\rm sil}=3.5\,\rm g\,cm^{-3}$, $\rho_{\rm gra}=2.24 \,\rm g\,cm^{-3}$ and $\rho_{\rm amc}=1.80 \,\rm g\,cm^{-3}$ for silicate, graphite and amorphous carbon, respectively. 

In this work, we consider two typical size distributions which are widely used to study the properties of interstellar dust. The first one is the MRN power-law size distribution: $dn_i/da = C_i a^{-\alpha_i}$, where $a$ is the grain radius in the range of $[a_{\rm min},a_{\rm max}]$, $\alpha$ is the power index, $dn_i$ is the number density of the dust grain of species $i$ with radius in the range of $[a,a+da]$, $C_i$ is a normalization constant related to the mass fraction of species $i$, given by:
\begin{equation}
C_i=\frac{m_i}{\int da (4 \pi /3) a^3 \rho_i a^{-\alpha_i}},
\end{equation}
where $\rho_i$ is the material density of species $i$, $m_i$ is the mass fraction of species $i$ expressed as percentages. Unfortunately, the detailed elements abundance of SN 2010jl is unknown. The host galaxy of SN 2010jl, UGC 5189A, is a low metallicity galaxy, with $Z\lesssim0.3\,Z_{\odot}$ \citep{sto11}. Therefore, the mass fraction of different species is initially set to be adjustable, which will be discussed in Section \ref{sec:mass_ratio}.

Another size distribution function used in this work is an exponentially cutoff power-law form: $dn_{i}/da=C'_{i}a^{-\alpha_i}{\rm exp}^{-a/a_{c,\,i}}$ \citep[KMH]{kim94},  where the $a_{c,\,i}$ is the exponentially cutoff index.
The KMH formula modifies the power-law function and makes the size distribution decrease exponentially with a smooth cutoff at the large-size side. For the KMH size distribution, the normalization constant $C'_i$ would be written as: 
\begin{equation}
C'_i=\frac{m_i}{\int da (4 \pi /3) a^3 \rho_i a^{-\alpha_i} e^{-a/a_{c,\,i}}}.
\end{equation}

Given the composition and dust size distribution, the total extinction at wavelength $\lambda$ is calculated by \citep{gao15,wan14}
\begin{equation}\label{ext}
\frac{A_{\lambda}}{N_{\rm H}}=(2.5\,{\rm log}e)\sum_i \int da  \frac{1}{n_{\rm H}}\frac{dn_i}{da}C_{{\rm ext}, i}(a,\lambda),
\end{equation}
where $N_{\rm H}$ ($n_{\rm H}$) is the column (number) density of hydrogen. With the assumption of sphere dust grains, $C_{{\rm ext}, i}(a,\lambda)$ is the extinction cross section of grain type $i$ of size $a$ at wavelength $\lambda$, calculated with Mie scattering theory \citep{mie08} from dielectric constants for given dust composition. In the case of graphite, we use 	``$\frac{1}{3}$-$\frac{2}{3}$ approximation'' to compute $C_{{\rm ext}, i}(a,\lambda)$ with the dielectric constant components perpendicular and parallel to the basal plane \citep{dra93}.

In Table \ref{tbl:models}, we summarize the details of models on dust components and size distributions. We fix $a_{\rm min}=0.001\,\mum$ for all the dust models. For MRN size distribution, the maximum cutoff radius $a_{\rm max}$ in Model A is a free parameter, while $a_{\rm max}$ in Model B is set to be $5.0\,\mum$ to explore the presence of large dust grains, which was proposed by \cite{gal14}. But for the KMH size distribution, we relax $a_{\rm max}=30.0\,\mum$ because $a_{\rm max}$ is coupling with the exponentially cutoff index $a_{c,\,i}$. These dust models are used to fit the observed extinction data of SN 2010jl from 66 - 239 days after peak brightness. 

It is widely used in many literature (e.g. \citealt{gal14,wei01}) that find a best-fit model to extinction through a reduced chi-square ($\chi^2/{\rm dof}$) minimization. Thus we minimize the $\chi^2/{\rm dof}$ to evaluate the goodness of fitting,
\begin{equation}\label{chi-square}
\chi^2/{\rm dof}=\frac{1}{N_{\rm obs}-N_{\rm para}}\sum_{j=1}^{N_{\rm obs}}\frac{[A_{\lambda,j}^{\rm obs}-A_{\lambda,j}^{\rm mod}]^2}{\sigma_j^2},
\end{equation}
where $A_{\lambda,j}^{\rm obs}$ and $A_{\lambda,j}^{\rm mod}$ are the observational extinctions and the modelled extinctions, respectively. 
$N_{\rm obs}$ is the number of observational data points, and $N_{\rm para}$ is the number of free parameters. $\sigma_j$ are the uncertainties of $A_{\lambda,j}^{\rm obs}$, which are purely determined by the signal-to-noise ratio (SNR) of hydrogen lines \citep{gal14}. It is worth mentioning here that the extinctions at H$\beta$ and P$\beta$ with higher SNR have much smaller uncertainties, i.e. more than 
an order of magnitude smaller than $\sigma_{\rm H\delta}$.

\begin{table*}
    \centering
    \caption{Dust models classified by grains components and size distributions. For all models here, the minimum cutoff radius $a_{\rm min}$ is fixed as $0.001\,\mum$. The parameter $m_{\rm sil}$ is the mass fraction of silicate, which is discussed in Section \ref{sec:mass_ratio}. \label{tbl:models}}
    \begin{tabular}{lccc}
    \hline
    \hline
    Dust models & $N_{\rm para}$ & Free paramerers & Constrains   \\
    \hline
    Model A: single component with MRN &2 & $a_{\rm max}$, $\alpha$ & single grain component \\
    Model B: silicate-graphite with MRN &2&$\alpha_{\rm sil},\,\alpha_{\rm gra}$&$a_{\rm max}=5.0\,\mum$, $m_{\rm sil}$ \\
    Model C: silicate-graphite with KMH &4 &$\alpha_{\rm sil}$, $\alpha_{\rm gra}$, $a_{c,\,\rm sil}$, $a_{c,\,\rm gra}$ & $a_{\rm max}=30.0\,\mum$, $m_{\rm sil}$  \\
    \hline
    \end{tabular}
\end{table*}

\subsection{Modelling the SEDs}

Generally, by assuming that the dust is optically thin, the total flux $F_{\rm \nu}(\lambda)$ is a combination of a blackbody emission of photosphere and the dust thermal emission, as calculated by \citep{gal14}
\begin{equation}\label{func_mass}
F_\nu(\lambda)=\frac{\pi B_\nu(\lambda,T_{\rm SN})R_{\rm SN}^2}{D^2}+ \frac{\sum M_{{\rm d},i}B_\nu(\lambda,T_{\rm d})\kappa_{i}(\lambda)}{D^2},
\end{equation}
where $B_{\nu}(\lambda,T)$ is the black-body Planck function at supernova temperature $T_{\rm SN}$ or dust temperature $T_{\rm d}$, with the assumption that all the dust grains have a uniform temperature $T_{\rm d}$; $R_{\rm SN}$ is the radius of supernova photosphere; $M_{{\rm d},i}$ is the dust mass of species $i$ around SN 2010jl; and $D$ is the luminosity distance to the observers \citep[i.e. $\rm 48.9\,Mpc$,][]{mae13}. The dust mass absorption coefficient $\kappa_i(\lambda)$ is a size-independent parameter and calculated with
\begin{equation}\label{func_abs}
\kappa_{i}(\lambda)=\frac{\int \kappa_{i}(\lambda,a) \frac{dn_i}{da}da}{\int \frac{dn_i}{da}da}.
\end{equation}

The size-dependent mass absorption coefficient $\kappa_{i}(\lambda,a)$ can be derived using $\kappa_{i}(\lambda,a)=3Q_{i}(\lambda,a)/4\rho_{i} a$, where $a$ is the radius of a spherical grain, $Q_{i}(\lambda,a)$ is the dust emission efficiency and can be calculated from the dust models described in Section \ref{subsec:ext-model} with the best-fit size distribution for the observed extinction curves of SN 2010jl. Finally, the size-independent parameter $\kappa_{i}(\lambda)$ can be derived by the convolution of $\kappa_{i}(\lambda,a)$ and $dn_{i}/da$.

\subsection{Mass Ratio between Silicate and Carbonaceous Grains}\label{sec:mass_ratio}

\begin{figure*}
	\centering
	\includegraphics[scale=0.6]{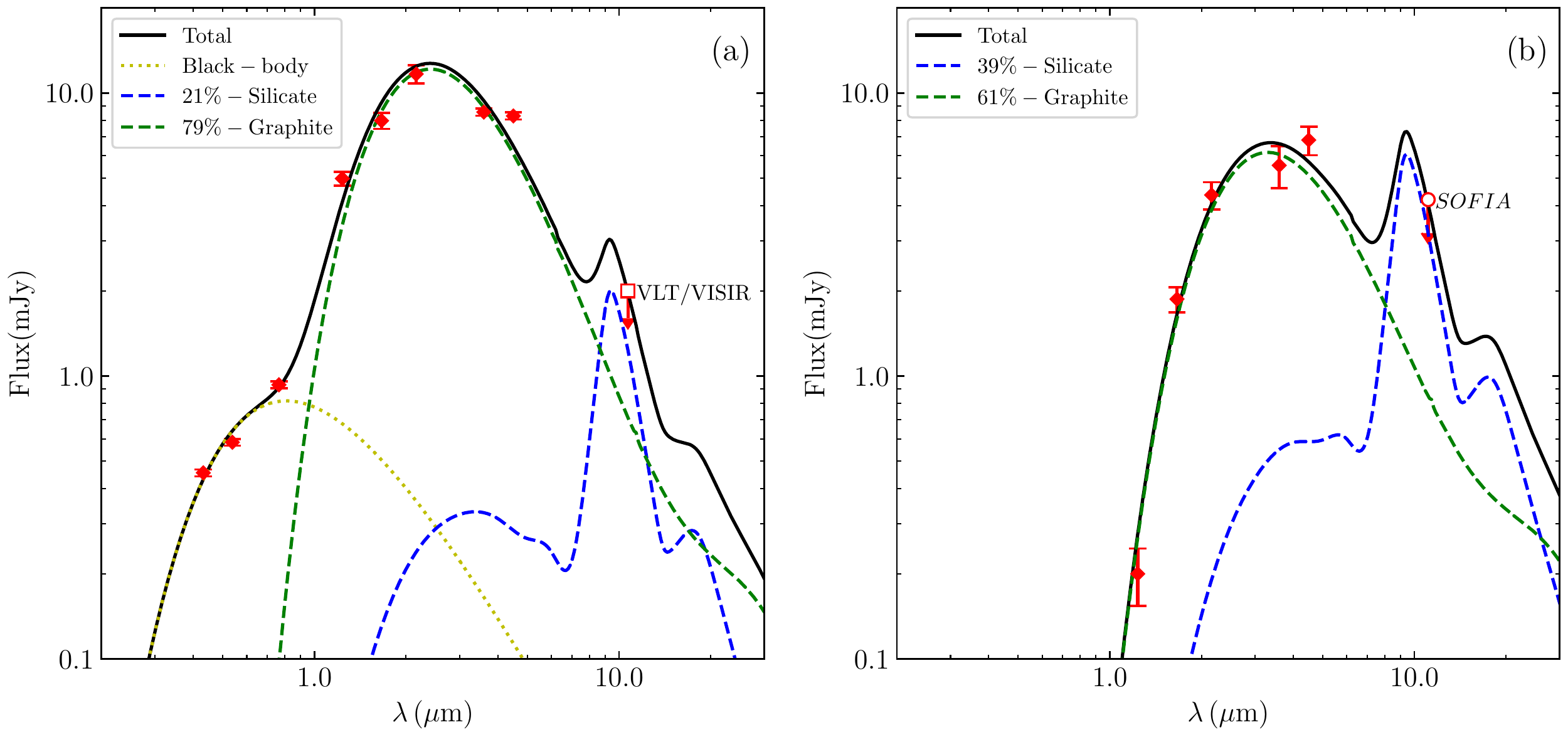}
	\caption{Results of constraining the silicate mass fraction. Panel (a): a two-component model with a supernova photosphere and the dust thermal emission from the mixture silicate and graphite fitting to the SED at 455 days. Panel (b): a dust thermal emission of mixed silicate and graphite fitting to the SED at 1294 days. The optical ($B,V,i'$) and NIR ($J,H,K_s$) data  taken from \protect\cite{fra14} and \emph{Spitzer} 3.6 and 4.5 $\mum$ data are all shown with red diamonds. The VLT/VISIR $10.7\,\mum$ and \emph{SOFIA} $11.1\,\mum$ flux upper limits are denoted with an open square or a circle with downward arrows, respectively. The predicted total SEDs are shown with black solid lines. The predicted contributions from the black-body, silicate and graphite grains are shown with yellow dotted lines, blue and green dashed lines, respectively. }
	\label{fig:mass-ratio}
\end{figure*}

\cite{wil15} roughly ruled out the presence of silicate around SN 2010jl based on modelling the \emph{Spitzer}/IRAC 3.6\,$\mum$ and 4.5\,$\mum$ emission and the flux upper limit at \emph{SOFIA} $11.1\,\mum$ with a single-size dust model of $a= 0.1\,\mum$. Noting that no optical/NIR photometry data included in their fitting. 

We adopt a mixed dust grains composed of astronomical silicate and graphite, which contribute to the dust thermal emission. Since the classical average size of dust is approximately 0.1\,$\mum$ in the MW \citep{lia01}, we simply assume all dust grains are spherical with the same radius of 0.1\,$\mum$. It has been found in previous works that dust grains formed in SN 2010jl at a temperature of $\sim 1000\,\rm K$ \citep[e.g.,][]{mae13,fra14}. With the optically-thin assumption and the dust temperature of $1000\,\rm K$, the emissivity (i.e. $\kappa B_{\nu}$) of silicate at $\lambda=11\mum$ is about 6 times higher than graphite. While the emissivity of silicate in the NIR bands is only a few percent of graphite. Therefore, the emission at $\sim\,11\,\mum$ would be dominated by silicate grains. If the mass percentage of silicate (i.e. $m_{\rm sil}$) is treated as a free parameter in fitting the measured SEDs together with the two flux upper limits around 11\,$\mum$ as mentioned in the last paragraph of Section \ref{sec:data_ir}, the best-fit $m_{\rm sil}$ would be an upper limit of silicate fraction.

At 455 days, the wavelength coverage of the photometry data is from optical to MIR bands, including ($B,V,i'$) and NIR bands ($J,H,K_s$) taken from \cite{fra14}, \emph{Spitzer} 3.6 and 4.5\,$\mum$ bands in Table \ref{tbl:mid-infrared} and 10.7\,$\mum$ of VLT/VISIR \citep{bev20}. A two-component fit consisting of the supernova photosphere and the dust emission is used at this epoch because of non-negligible contribution from the supernova photosphere. However, at 1294 days, there is no optical observation. We therefore only include the NIR ($J,H,K_s$) data at 1133 days from \cite{fra14}, \emph{Spitzer} 3.6 and 4.5\,$\mum$ data by averaging the fluxes at 988, 1358 and 1359 days, as well as \emph{SOFIA} $11.1\,\mum$ flux upper limit. A single-component with only dust emission is applied to fit the SED at 1294 days.

Figure \ref{fig:mass-ratio} shows the SEDs at 455 and 1294 days. From the best-fit results, we find that the estimated mass fractions of silicate are 21\% and 39\% for 455 and 1294 days, respectively. It still remains a subject of debate that whether the silicate grains exist around SN 2010jl. We only conservatively take into account the case of 455 days, which constrain the $m_{\rm sil}$ to a smaller value. Therefore, the mass percentage of silicate $m_{\rm sil}$ in Model B and C is set as 20\% in the subsequent calculations, as listed in Table \ref{tbl:ext-results}.

\section{Results}\label{sec:result}

\begin{table*}
    \centering
    \caption{Model parameters and best-fit results for the extinction curve of SN 2010jl at 239 days after peak brightness. We fix $a_{\rm min}=0.001\,\mum$ and $a_{\rm max}=a_{\rm max,sil}=a_{\rm max,gra}$ for all models.\label{tbl:ext-results}}
    \begin{tabular}{ccccccccc}
    \hline\hline
    Models & \multicolumn{2}{c}{dust} & $a_{\rm max}(\mu \rm m)$ & \multicolumn{2}{c}{$dn_i /da$}& $\chi^2/{\rm dof}$ &$\AV(\rm mag)$&$\RV$   \\

    \hline
    \multirow{3}{*}{Model A}&\multicolumn{2}{c}{100\% sil}    & 6.5 & \multicolumn{2}{c}{$a^{-3.6}$} &300.0 &0.67 &6.3 \\
    &\multicolumn{2}{c}{100\% gra} & 6.0 & \multicolumn{2}{c}{$a^{-3.6}$} & 281.5 & 0.66 & 5.7 \\
    &\multicolumn{2}{c}{100\% amc} & 5.6 & \multicolumn{2}{c}{$a^{-3.7}$} & 296.3 & 0.66 & 6.0 \\
    \hline
    Model B&20\% sil&80\% gra & 5.0 & $a^{-4.37}$&$a^{-3.54}$ & 273.0 &0.66 & 5.9\\
    \hline
    Model C&20\% sil&80\% gra & 30.0 & $a^{-2.23}e^{-a/0.03}$ &$a^{-1.77}e^{-a/3.39}$& 27.9 & 0.57 & 3.0\\

    \hline
    \end{tabular}

\end{table*}

\subsection{Modelling the Extinction Curve at 239 Days \label{subsec:result_239}}

\begin{figure*}
\centering
\includegraphics[scale=0.7]{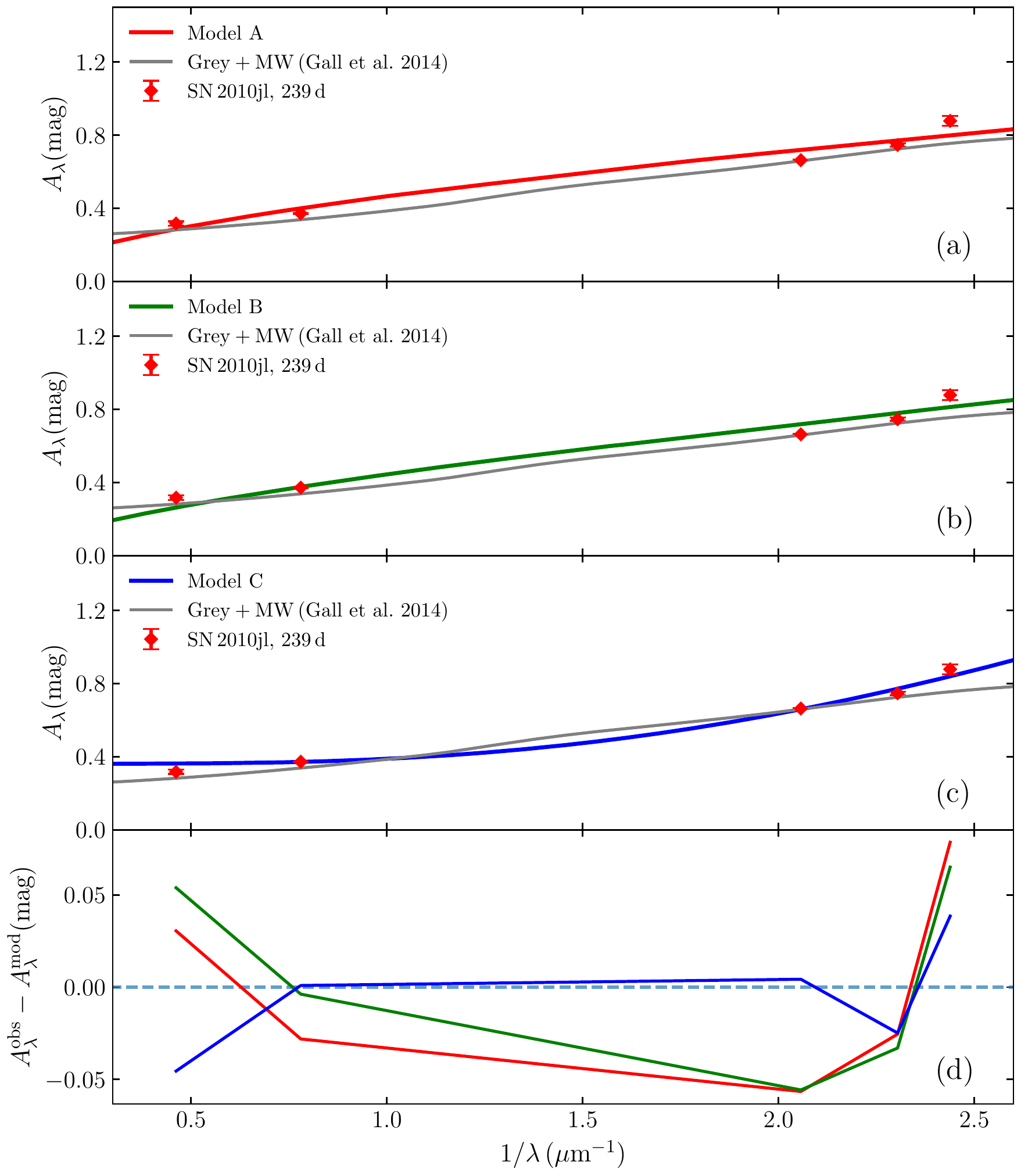}
\caption{Comparisons of the observed SN 2010jl extinction at 239 days with the best-fit extinction curves (colored solid lines) calculated with (a) the amorphous carbon model with MRN size distribution in Model A, (b) silicate-graphite mixed model with different MRN size distributions in Model B, and (c) silicate-graphite model with different KMH size distributions in Model C. The grey solid line is the combined extinction curve proposed by \protect\cite{gal14} with 40$\%$ grey extinction and 60$\%$ the MW extinction in $V$ band. Residuals between the observed extinction and modelled extinction are shown in panel (d). The mass ratio of silicate to graphite for Model B and C here is set as $20\% : 80 \%$.}
\label{fig:abc}
\end{figure*}

As shown in Figure \ref{fig:ccm}, the extinction laws of SN 2010jl change little with time except in the $\rm Br\gamma$ band. We first fit the extinction curve at 239 days using three types of model described in Table \ref{tbl:models} to test the feasibility of our models. The best modelled extinction curves in Model A, B and C are displayed in Figure \ref{fig:abc}, and the corresponding parameters are listed in Table \ref{tbl:ext-results}.

In case of Model A, the extinction curve of SN 2010jl at 239 days is fitted by using a single grain component, i.e. silicate, graphite or amorphous carbon (see the first three rows in Table \ref{tbl:ext-results}), respectively. The derived $\RV$ values are 6.3, 5.7 and 6.0, which are very close to $\RV\approx6.4$ reported by \cite{gal14}. It is well known that the average extinction law of the MW can be reproduced with the silicate-graphite dust model with a standard MRN size distribution \citep{mat77}, i.e. $a^{-3.5}$ with $a_{\rm min}=0.005\,\mum$ and $a_{\rm max}=0.25\,\mum$. \cite{gal14} used amorphous carbon model with MRN size distribution to fit their extinction curves of SN 2010jl, and obtained the best-fit parameters of $a_{\rm max}=4.2\,\mum, \alpha =3.6$. With only amorphous carbon dust in Model A, our best-fit results are achieved with $a_{\rm max}=5.6\,\mum, \alpha =3.7$, similar to the results of \cite{gal14}. The best-fit values of $a_{\rm max}$ in Model A and that of \cite{gal14} are much larger than that of MRN for the MW, i.e. $a_{\rm max}=0.25\,\mum$. It suggests that a large fraction of micron-size dust grains is needed to well reproduce the observed extinction curves of SN 2010jl. These two modelled extinction curves with individual component of amorphous carbon are shown in the panel (a) of Figure \ref{fig:abc}. However, it is clear that this single component model with MRN size distribution does not well reproduce the unusual extinction law of SN 2010jl. Especially, the modelled extinction value at the 0.4101\,$\mum$ (H$\delta$) band is much smaller than the observed one, suggesting that a portion of small-size dust grains are also required to explain the steep increasing extinction in short-wavelength bands.

We also fit the extinction curve of SN 2010jl at 239 days with the silicate-graphite mixed models described in Model B and C, by considering different MRN and KMH distributions for different components, i.e. $\alpha_{\rm sil}\neq\alpha_{\rm gra}$, $a_{\rm c,\,sil}\neq a_{\rm c,\,gra}$. In these cases, only graphite is considered as the carbonaceous component. According to the results from single-component dust models in Model A, it is reasonable to fix $a_{\rm max}=5.0\,\mum$ for both silicate and graphite in Model B. According to the discussion about the silicate mass percentage in Section \ref{sec:mass_ratio}, the calculations in Model B and C are processed with a fixed $m_{\rm sil}$ of 20\%.

From the best-fit results in Table \ref{tbl:ext-results} and Figure \ref{fig:abc}, the Model B gives best-fit results of $\RV\approx5.9$ and $\AV \approx 0.66$ mag, which are very close to the best-fit results of Model A. The Model C provides a smallest $\chi^2/{\rm dof}$ among the three models, leading to $\AV \approx 0.57$ mag, which is consistent with 0.6 mag derived by \cite{gal14}, while $\RV \approx 3.0$ is significantly smaller than 6.4 derived by \cite{gal14}. The residuals between our modelled extinction and observed extinction can be seen in Figure \ref{fig:abc}(d), it is noted that the residuals of Model C are overall smaller than the residuals of Model A and B. More importantly, Model C can better reproduce the observed extinction excess at H$\delta$, the bluest waveband of observation that can best constrain the small dust grains. By comparing the $\chi^2/{\rm dof}$ values, it shows that both of Model A and B provide equally good fitting (i.e. $\chi^2/{\rm dof}\sim 300$), we suggest that Model C with KMH size distribution ($\chi^2/{\rm dof}=27.9$) is more suitable to explain the extinction curves of SN 2010jl.

\begin{figure*}
\centering
\includegraphics[scale=0.7]{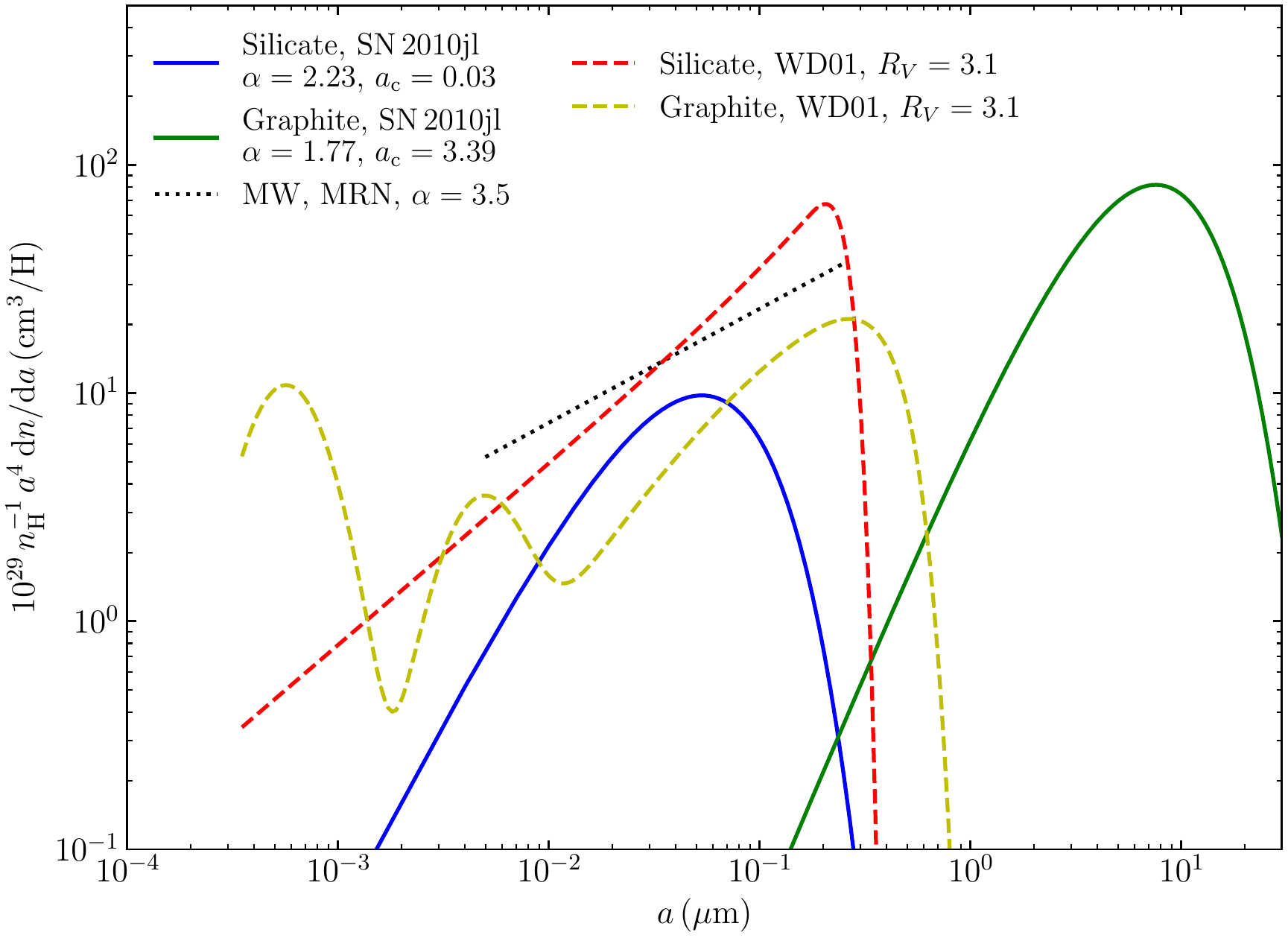}
\caption{Comparison of size distributions of silicate (blue solid line) and graphite (green solid line) with $m_{\rm sil}=20\%$ and $m_{\rm gra}=80\%$ of Model C. The WD01 \citep{wei01} size distributions of silicate and graphite for the average extinction of Milky Way with $\RV\,=\,3.1$ are shown by dashed lines (red for silicate and yellow for graphite). The MRN size distribution with $\alpha=3.5$ \citep{mat77} is also shown by black dotted line. \label{fig:size-dist}}
\end{figure*}

Figure \ref{fig:size-dist} shows the size distributions of silicate and graphite derived from the best-fit results of Model C with $m_{\rm sil}=20\%$. The size distribution of silicate grains shows a sharp decrease and skews towards the smaller dust grains, which may account for the steep extinction curve on the side of short wavelength range. In contrast, the peak value of size distribution of graphite grains extends to around $a\sim 3.0\,\mum$, which implies the presence of large-size graphite dust grains. Recently, \cite{zha20} also found that the silicate dust grains in supernova remnants are on average smaller than the carbonaceous grains. The different size distributions between the silicate and graphite grains may originate from the formation process of different types of dust grains in the environment of supernovae explosion \citep{bev17}.

\subsection{Modelling the Extinction Curves of SN 2010jl at 66 -196 Days\label{subsec:result_all}}

\begin{table*}
\centering
    \caption{Best-fit results for modelling the extinction curves of SN 2010jl from 66 to 239 days, using Model C with $m_{\rm sil}=20\%$ and $m_{\rm gra}=80\%$. We fix $a_{\rm min}=0.001\mum$ and $a_{\rm max}=a_{\rm max,\,sil}=a_{\rm max,\,gra}=30.0\mum$ for models in Model C.\label{tbl:ext-results-all}}
    \begin{tabular}{ccccccccc}
    \hline\hline
    Epoch & $dn_{\rm sil} /da$& $dn_{\rm gra} /da$ &  $\chi^2/{\rm dof}$ & $A_U$ & $A_B$ & $\AV$ & $\RV$ & $N_{\rm d}$\\
    (days)& & & & (mag) & (mag) &(mag) & & $(\times 10^{-4}\,\rm g\,cm^{-2})$\\
    \hline
    66  & $a^{-2.10}e^{-a/0.02}$ & $a^{-2.39}e^{-a/4.60}$ & 31.5& 0.32 & 0.26 & 0.22 & 5.5 & 0.87\\
    104 & $a^{-1.85}e^{-a/0.02}$ & $a^{-1.95}e^{-a/3.65}$ & 92.7& 0.53 & 0.41 & 0.34 & 4.8 & 1.40\\
    121 & $a^{-1.85}e^{-a/0.02}$ & $a^{-1.43}e^{-a/4.00}$ & 9.8& 0.63 & 0.46 & 0.34 & 2.8 & 1.92\\
    140 & $a^{-2.72}e^{-a/0.03}$ &	 $a^{-1.56}e^{-a/3.85}$ & 66.2& 0.75 & 0.55 & 0.41 & 2.9 & 1.98\\
    158 & $a^{-2.61}e^{-a/0.03}$ & $a^{-1.76}e^{-a/3.90}$ & 91.2& 0.82 & 0.61 & 0.46 & 3.1 & 1.93\\
    196 & $a^{-2.21}e^{-a/0.02}$ & $a^{-1.70}e^{-a/5.80}$ & 85.9& 1.05 & 0.76 & 0.56 & 2.8 & 3.69\\
    239 & $a^{-2.23}e^{-a/0.03}$ & $a^{-1.77}e^{-a/3.39}$ & 27.9& 1.01 & 0.76 & 0.57 & 3.0 & 2.06\\
    \hline
    \end{tabular}
\end{table*}

\begin{figure*}
\centering
\includegraphics[scale=0.6]{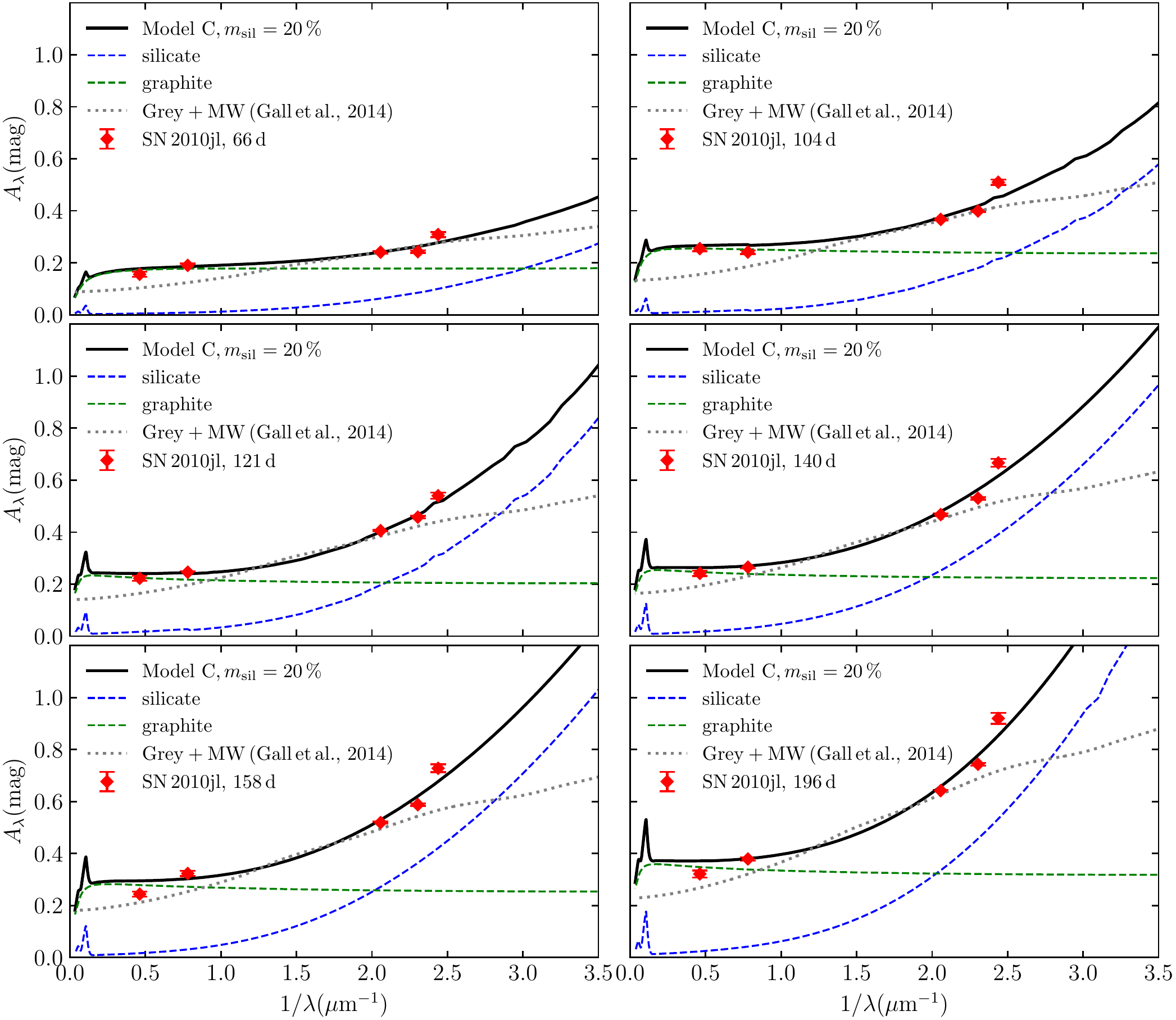}
\caption{Comparisons of the modelled extinction curves (black solid lines) derived by Model C with the observed extinction (red diamonds) from 66 days to 196 days. The modelled extinction curves contributed by silicate and graphite are also plotted by blue and green dashed lines, respectively. Grey+MW extinction curves described in \protect\cite{gal14} are also plotted with grey dotted lines.}
\label{fig:ext-all}
\end{figure*}

As the description in the above subsection, the silicate-graphite dust model characterized with $m_{\rm sil}=20\%$ and KMH size distribution could produce a smallest $\chi^2/{\rm dof}$ among the three models, when fitting the extinction curve of SN 2010jl at 239 days. In order to investigate the time variation of size distributions of dust grains around SN 2010jl, we then fit all the extinction curves from 66 to 239 days derived by \cite{gal14} using Model C.

In Table \ref{tbl:ext-results-all}, we show the best-fit parameters and results obtained from fitting the extinction curves from 66 to 239 days using Model C with $m_{\rm sil}=20\%$. The best-fit results give that the size distributions of silicate have similar cutoff values $a_{\rm c,\,sil} \approx 0.02\,-\,0.03\,\mum$. Such small values of $a_{\rm c,\,sil}$ suggest that the best model of Model C is rich in small-size silicate grains, which contribute to the steep increase of extinction curves in short wavelength bands. The power law index $\alpha_{\rm sil}$ of silicate shows no obvious change with time.
On the contrary, large values of the cutoff size for graphite $a_{\rm c,\,gra}\approx 3-5\,\mum$ are obtained, which suggests that there are a lot of large-size graphite grains in the best-fit models, which contribute to the flat and grey extinction in the optical and NIR bands. There is also no time-dependent variation for power-law indices $\alpha_{\rm gra}$ in the size distribution of graphite. 

From Table \ref{tbl:ext-results-all}, the best-fit results show that the modelled extinction in $V$ band $\AV$ increases with time, from 0.22 mag at 66 days to 0.57 mag at 239 days continuously, which is consistent with what obtained by \cite{gal14}. The modelled total-to-selective ratios $\RV$ are around 2.8 - 3.1 for almost all epochs except $\RV \sim 5.0$ at early 66 and 104 days. All the comparisons between observed and modelled extinction curves are shown in Figure \ref{fig:ext-all}. The best-fit models show that the graphite grains contribute to a flat or grey extinction to the total extinction curves of SN 2010jl at all epochs, while the silicate grains contribute to the steep rise in the short wavelength bands. Consequently, the silicate-graphite dust model of Model C in this work provides more reasonable explanation on the observed extinction curves of SN 2010jl than the grey+MW/SMC extinction model provided by \cite{gal14}.

\section{Discussion}\label{sec:dis}

\subsection{Dust Mass}

\begin{figure*}
\centering
\includegraphics[scale=0.8]{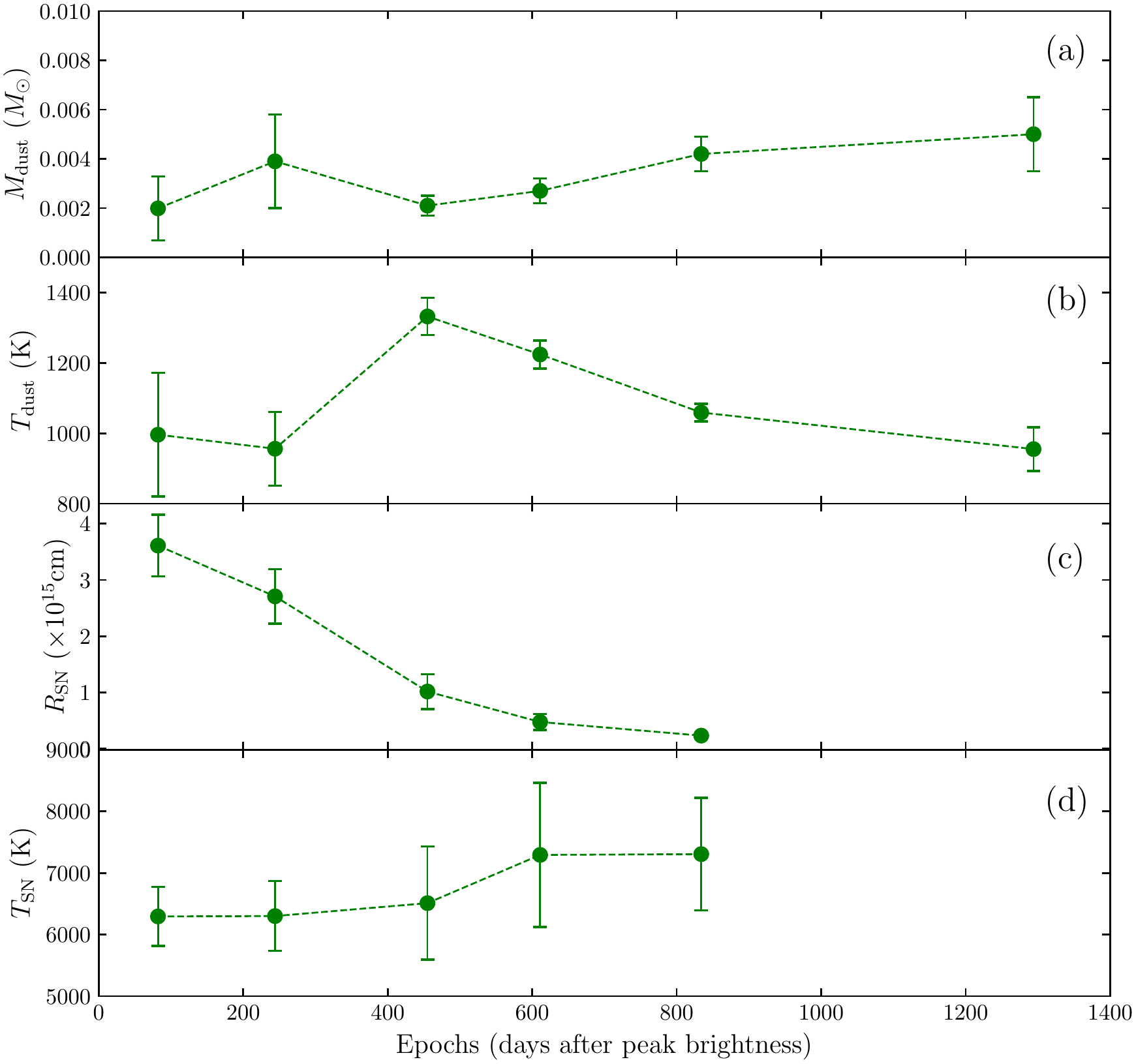}
\caption{The dust masses (a), the dust temperatures (b), the black-body radius (c) and the black-body temperatures (d) of SN 2010jl as a function of time after the peak brightness, which are obtained by fitting the SEDs using modelled mass absorption coefficient $\kappa_i(\lambda)$.}
\label{fig:mass-dist}
\end{figure*}

\begin{table*}
    \centering
    \caption{Parameters of fitting SEDs of SN 2010jl. Figures in parentheses indicate the uncertainties of estimated parameters. \label{tbl:dust-mass}}
    \begin{tabular}{ccccccc}
    \hline\hline
    Epoch (days)           &    82 &   244  &   455 &   611 &   834 &   1294  \\
    \hline
    $T_{\rm SN}$\,(K) & 6295(480) & 6301(567)&  6511(920)& 7294(1170)& 7307(913)& - \\
    $R_{\rm SN}$\,($\times 10^{15}\,$cm) &3.61(0.55) &2.71(0.48)  &1.02(0.31) & 0.47(0.14) & 0.23(0.05) & -\\
    $T_{\rm d}$ (K) &   996(176) &   956(105)  &   1332(53) &   1224(40) &  1059(25) &  955(62)\\
    $M_{\rm d}\,(M_\odot)$ & 0.0020(0.0013) & 0.0038(0.0019)  & 0.0021(0.0004) & 0.0027(0.0005) & 0.0042(0.0007) & 0.005(0.0015) \\
    \hline
    \end{tabular}
\end{table*}

\begin{figure*}
\centering
\includegraphics[scale=0.8]{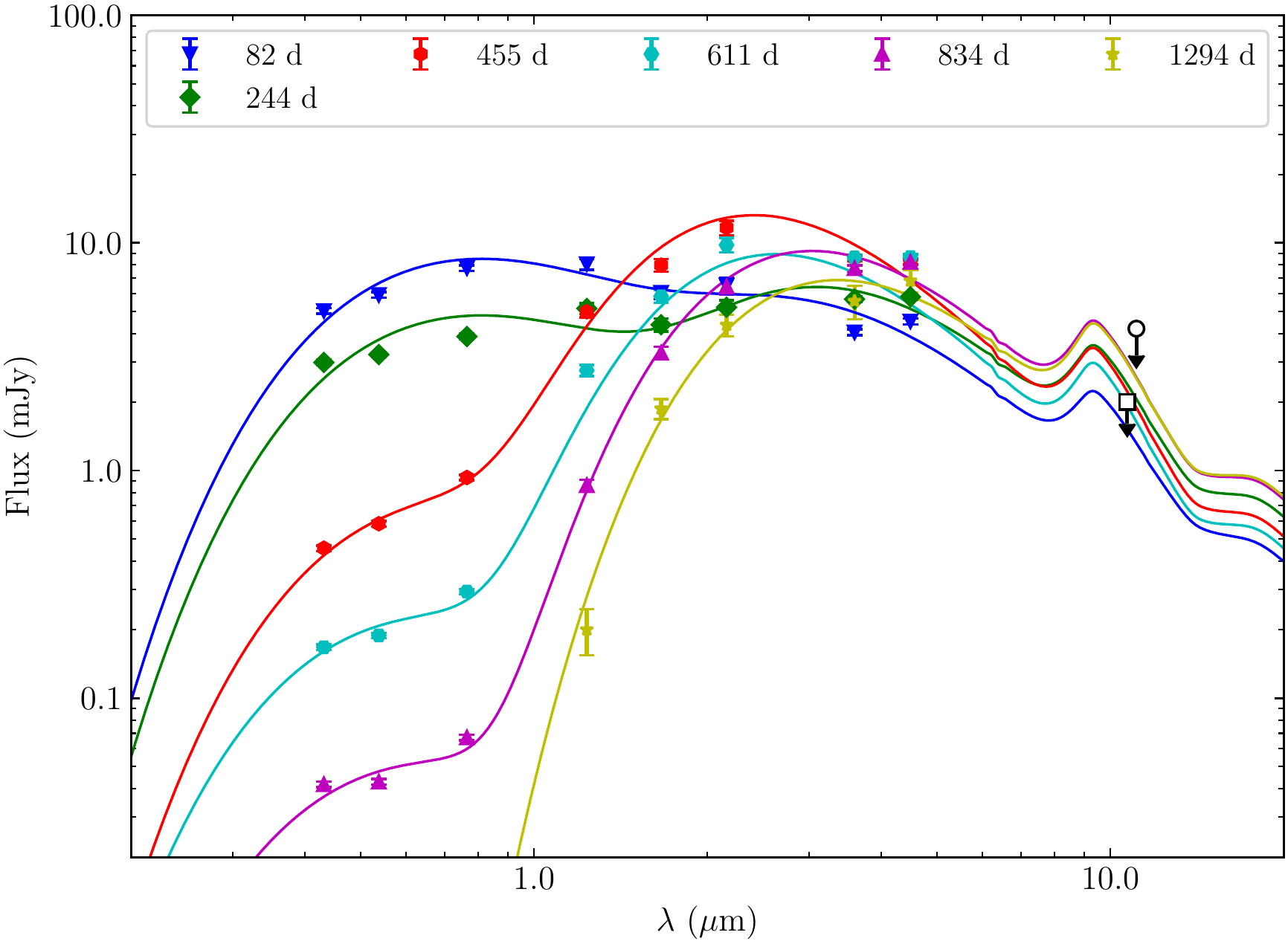}
\caption{Best-fit to the SEDs of SN 2010jl using Equation \ref{func_mass} with modelled $\kappa_{i}(\lambda)$ at different epochs. The flux upper limits at 10.7\,$\mum$ and 11.1\,$\mum$ observed by VLT/VISIR \citep{bev20} at 455 days and $SOFIA$ \citep{wil15} at 1294 days are plotted by an open square and circle with downward arrows, respectively. Note that the two upper-limit points are not adopted in fitting the SEDs of SN 2010jl. \label{fig:sed-fit}}
\end{figure*}

The IR excesses of SN 2010jl were detected in the very early stage after explosion (about three months), and still very bright in the next few years. \cite{sar18} adopted a two-component model to fit the SEDs of SN 2010jl from optical bands to MIR bands, and suggested that the hot component from photosphere of SN 2010jl contributed a little to the MIR emission. Particularly, after 244 days, the dust emission dominates the SEDs of SN 2010jl in the NIR and \emph{Spitzer} MIR bands \citep{sar18}. Using a standard MRN size distribution of $\propto\,a^{-3.5}$ within $[a_{\rm min}=0.005\,\mum, a_{\rm max}=0.05\,\mum]$, \cite{and11} estimated a dust mass between 0.03 and 0.35\,$M_\odot$, by fitting the photometric data at 90 days in the 3.6\,$\mum$, 4.5\,$\mum$ bands of \emph{Spitzer}/IRAC and the JHK bands of WIYN. \cite{mae13} derived a dust mass of $\sim (7.5-8.5)\times 10^{-4}\,M_\odot$ from an optical to NIR spectrum around 1.5 years after the explosion.

By using Equation \ref{func_mass}, the optical ($B,V,i'$) and NIR ($J,H,K_s$) data taken from \cite{fra14} and \emph{Spitzer}/IRAC data listed in Table \ref{tbl:mid-infrared} are fitted to find the dust masses and temperatures. The mass absorption coefficient $\kappa$ (in unit of $\rm cm^2\,g^{-1}$), is calculated with Equation \ref{func_abs} by using the modelled size distributions shown in Table \ref{tbl:ext-results-all}. For the SED at 82 days, we use the modelled size distribution corresponding to best-fit results from extinction Model C at 66 days, as listed in Table \ref{tbl:ext-results-all}. Because the size distributions of both silicate and graphite grains do not change a lot within 239 days (see Table \ref{tbl:ext-results-all}), we then utilize the composition and size distributions of Model C at 239 days to do the estimation of dust masses and temperatures at epochs later than 244 days.

The derived dust masses and temperatures using modelled $\kappa_i(\lambda)$ are listed in Table \ref{tbl:dust-mass}. Figure \ref{fig:mass-dist} presents the dust masses, dust temperatures, supernova black-body radii and temperatures as a function of time. With $m_{\rm sil}=20\%$, the dust models produce the silicate feature around 9.7\,$\mum$ at all epochs. Although the upper-limit fluxes of VLT/VISIR 10.7\,$\mum$ and \emph{SOFIA} 11.1 $\mum$ are not adopted in fitting the SEDs, the predicted fluxes do not show much excessive emission at 10.7\,$\mum$ than the flux upper limit of VLT/VISIR at all epochs, and are all smaller than the flux upper limit of \emph{SOFIA}, as clearly seen in Figure \ref{fig:sed-fit}.
From Table \ref{tbl:dust-mass} and the panel (a) of Figure \ref{fig:mass-dist}, it is clear that the dust mass increases continuously with time until 1294 days when the dust mass reaches about 0.005\,$M_\odot$ and the dust temperature decreases to about 955\,K. \cite{bev20} also found that there has been continuously dust forming in SN 2010jl from the photometric and spectral observations. Additionally, the dust temperatures show a short-term increasing in the first 500 days after peak brightness, then recede afterwards. The estimates of radii and temperatures of photosphere are consistent with the results of previous works \citep{sar18,gal14,fra14}. 

Table \ref{tbl:ext-results-all} lists the dust mass column density $N_{\rm d}$ (in units of $\rm g\,cm^{-2}$) calculated with the best-fit models for the extinction curves of SN 2010jl at all epochs. \cite{zha18} estimated the dust mass of Monoceros supernova remnant by using the equation of $M_{\rm d} = N_{\rm d}\times A_{\rm eff}$, where $A_{\rm eff}$ is the effective surface area of the supernova remnant.
The effective surface area can be calculated with  $A_{\rm eff}=\pi R^{2}_{\rm CDS}$, where $R_{\rm CDS}$ is the radius of cold dense shell, where dust grains were formed. Using a two-blackbody model, \cite{sar18} calculated the infrared radius of SN 2010jl at 244 days to be $2.2\,\times\,10^{16}\,\rm cm$, which is consistent with the radius of blackbody derived by \cite{gal14}. With $N_{\rm d}\approx 2.06\,\times10^{-4}\,\rm g\,cm^{-2}$ at 239 days, the yielded dust mass is around $2.0\,\times \,10^{-4}\,M_\odot$, that is about one order of magnitude smaller than the results from modelling the SED at 244 days. Note that the radius of the blackbody is just a lower limit of the size of dust shell \citep{fox13,sar18}, thus the dust mass calculated with this radius should be regarded as the lower limit and smaller than that derived from the infrared emission. Besides, an inhomogeneous and clumpy structure of dust distribution around SN 2010jl could bring deviation to the estimation of dust mass \citep{nie18}. 

In Figure \ref{fig:sed-fit}, the observed 4.5\,$\mum$ fluxes are slightly higher than the predicted ones at all epochs. \cite{kot09} reported that a type IIP SN 2004et was strongly affected by CO. So the CO emission may contaminate the \emph{Spitzer} 4.5\,$\mum$ flux of SN 2010jl. By adding the optical and NIR data into SED fitting, it will minimize the contamination from CO. With the expansion of the supernova, the dust temperature decreases so that the emission shift to longer wavelength which is not observed. The emission at 3.6 and 4.5\,$\mum$ only accounts for the warm dust while not for the relatively cold dust. The lack of longer wavelength observations may under-estimate the dust mass.

\subsection{Dust Components in SN 2010jl}

Stellar environments are mainly rich in either carbon or oxygen, predominantly carbonaceous dust or silicate will form. Current observation of SN 2010jl can not resolve the dust species. Thus a widely-used standard composition of silicate and graphite from \cite{dra84} is used to model the extinction of SN 2010jl. We also use amorphous carbon from \cite{zub96} or \cite{han88} instead of graphite, a mixture of silicate and amorphous carbon could bring a equally good reproduction of the extinction of SN 2010jl. Actually, different carbonaceous species are interchangeable in fitting to extinction curves of SN 2010jl. In addition, it is still controversial whether the silicate grains exist around SN 2010jl. 

Without any MIR data in their fitting, \cite{mae13} suggested that the main species of dust around SN 2010jl are carbonaceous grains. The observations by both VLT/VISIR \citep{bev20} at 510 days and \emph{SOFIA} \citep{wil15} at 1294 days did not detect any obvious fluxes of SN 2010jl at 10.7 and 11.1\,$\mum$, respectively. Although both \cite{gal14} and \cite{fra14} argued that silicate grains are less likely to exist within $R_{\rm CDS}$ due to the evaporation by the blast shock, it still cannot rule out the presence of silicate grains completely, because the calculated evaporation radius of silicate is larger than that of graphite by only a factor of around two \citep{gal14,fra14}. It will occur under the thermodynamical equilibrium between the solid and the gas phases \citep{len95}, but the efficiency of evaporation is quite uncertain. There is also a plausible explanation that the dust grains formed in small dense clumps, in which the cooling is more efficient \citep{kot09}. Furthermore, \cite{sar18} yields a significant amount of silicate mass around SN 2010jl through the stoichiometric calculation.
They also pointed out that the presence of silicate is still possible if the size of dust grains is as large as up to $5\,\mum$ or there is an optically thick dust shell. The self-absorption in the optically thick clumps may also cause the absence of silicate features \citep{dwe19}. In addition, silicate grains are reported to dominate the dust grains in some young SN remnants \citep{bev17}. \cite{fab11} also found a mixture with 80\% silicate and 20\% amorphous carbon can lead to a good fit to the SEDs of Type II-P supernova SN 2004et.

Consequently, it is possible that the  composition around SN 2010jl consists of a mixture of silicate and carbonaceous grains. By using the two flux upper limits from \cite{wil15} and \cite{bev20}, a conservative estimate gives the mass percentage of silicate grains $m_{\rm sil}\approx20\%$, as discussed in Section \ref{sec:mass_ratio}. Interestingly, the silicate mass fraction of $20\%$ gives well fits for both the extinction curves and SEDs.

\subsection{The Extinction Law of SN 2010jl}

In general, the total-to-selective extintion ratio $\RV$ is an empirical parameter to characterize the extinction law of the MW. The value of $\RV$ highly depends on the properties of dust grains, i.e. importantly the size distribution. A large value of $\RV$ results from the larger-size dust grains \citep{dra03}, thus the extinction curve of a very large $\RV$ may be very flat or ``grey". The average extinction law for the diffuse regions in the local MW agrees with $\RV\approx3.1$ \citep{dra03}.

A few previous studies show that some Type IIn supernovae have larger value of $\RV$.  
For example, \cite{nie18} found that $\RV$ of the type IIn supernova SN 2005ip increased and became stable in the range of 4.5 to 8.0 after 100 days.
However, it is still controversial whether such large value of $\RV$ is universal for the type IIn supernovae. In this work, the best-fit Model C with $m_{\rm sil}=20\%$ and $m_{\rm gra}=80\%$ yields $\RV\approx 2.8-3.1$, which is much smaller than $\RV\approx 6.4$ derived by \cite{gal14}, but close to the average $\RV\approx3.1$ of the MW. Although the values of $\RV$ are very close, the actual shape of extinction curves of SN 2010jl are distinctly different from that of the MW (see the extinction curve of $\RV=3.1$ in Figure \ref{fig:ccm}).
Therefore, it suggests that a single parameter of $\RV$ is insufficient to completely represent the extinction curve of SN 2010jl and the detailed dust properties around SN 2010jl. The empirical single parameter extinction curves of \citetalias{car89} or \citetalias{fit99} is not suitable and sufficient to explain the extinction curves of Type IIn supernovae or the extinction laws outside the MW \citep{gao15,lia01,gao19}. From the results in Table \ref{tbl:ext-results-all}, there is no obvious time-dependent variation of $\RV$ after 100 days for the dust around SN 2010jl.

From the best-fit results in Table \ref{tbl:ext-results-all}, Model C with silicate mass fraction of 20\% predicts that the extinctions in the $U$ and $B$ bands (i.e. $A_U$ and $A_B$) increase from 0.32\,mag and 0.26\,mag at 66 days to 1.05\,mag and 0.76\,mag at 196 days, respectively. The corresponding increasing rates per 100 days are 0.56\,mag in $A_U$ and 0.38\,mag in $A_B$. Although the evolution of light curves shows a faster decreasing rate in blue bands from the peak brightness to 90 days \citep{zha12}, i.e. 1.17\,mag per 100 days in $U$ band and 0.96\,mag per 100 days in $B$ band, the extinction growth would not contribute to the declining of light curves significantly.

\section{Conclusions}\label{sec:con}

In this work, we investigate the properties of dust in Type IIn supernova SN 2010jl by fitting the observed extinction curves derived by \cite{gal14}.
With the mixture dust models consisting of silicate and graphite grains with different exponentially cutoff power-law size distributions, the extinction curves of SN 2010jl from 66 to 239 days after peak brightness are well reproduced. The main results and conclusions are as follows:

\begin{enumerate}
	\item{The best-fit dust models include both silicate and graphite grains, and it is proposed that the mass percentage of silicate grains around 2010jl is constrained to be 20\%. }
	\item{Dust models with different KMH size distributions for the silicate and graphite grains well reproduce the observed extinction curves of SN 2010jl.}
	\item{The best-fit results show that the visual extinction $A_V$ of SN 2010jl increases with time, from 0.22\,mag at 66 days to 0.57\,mag at 239 days.}
	\item{A single parameter of $\RV$ is insufficient to completely represent the extinction curves of SN 2010jl. The derived $\RV$ values of SN 2010jl are found to be 2.8 - 3.1, which are close to the average value $\RV\,\approx\,3.1$ of the MW and much smaller than the value of $\RV\approx6.4$ derived by \cite{gal14}.}
	\item{By fitting the optical to MIR photometry of SN 2010jl with the mass absorption coefficients calculated from the best-fit dust models, the dust mass around SN 2010jl is found to be increasing with time, to $0.005\,M_\odot$ at around 1300 days, which is consistent with the results in literature \citep{gal14,sar18}.}
\end{enumerate}

\section*{Acknowledgements}

We are very grateful to the referees for their helpful comments. We thank Prof. Li, A. and Prof. Wang, D. for their useful suggestions/comments. This work is supported by NSFC Projects 12133002, 11533002, U1631104 and U2031209, National Key R\&D Program of China No. 2019YFA0405503 and CMS-CSST-2021-A09. JL acknowledges support by the China Scholarship Council (No.\,201706040153) for the study at University of Massachusetts, Amherst.

\section*{Data Availability}
The data underlying this article will be shared on reasonable request to the corresponding author.



\bibliographystyle{mnras}
\bibliography{sn2010jl} 





\bsp	
\label{lastpage}
\end{document}